
\documentstyle{amsppt}
\magnification 1200
\NoBlackBoxes
\document

\def\Ml{M_{\lambda}}
\def\d{\partial}
\def\Uloc{U(\g)_{\text{reg}}}
\def\hatglN{\frak g_0}
\def\g{\frak g_0}
\def\glN{\frak{gl}_N}
\def\vlk{v_{\lambda,k}}
\def\l{\lambda}
\def\Mlk{M_{\lambda,k}}
\def\Vm{V_{\mu}}
\def\pslk{\psi_{\lambda,k}}
\def\bz{\bold z}
\def\deg{\text{deg}}
\def\ord{\text{ord}}
\def\sdeg{\text{sdeg}}
\def\Ucr{U_{\text{crit}}}
\def\hatslN{\widehat{\frak{sl}_N}}
\def\SLN{\frak{sl}_N}

\centerline {\bf Quantum integrable systems
and representations of Lie algebras}
\vskip .15 in
\centerline{\bf Pavel I. Etingof}
\vskip .1in
\centerline{Department of Mathematics}
\centerline{Yale University}
\centerline{New Haven, CT 06520}
\centerline{e-mail etingof\@ pascal.math.yale.edu}
\vskip .1in

\centerline{\bf Introduction}

A quantum many particle system on the line with an interaction
potential $U(x)$ is defined by the Hamiltonian
$$
H=\sum_{j=1}^N\frac{\d^2}{\d x_j^2}+K\sum_{1\le i<j\le
N}U(x_i-x_j),\ K\in\Bbb R.
\tag 1
$$
Such systems were first considered by Calogero for $U(x)=x^{-2}$ \cite{Ca}
and Sutherland for $U(x)=\text{sinh}(x)^{-2}$ \cite{Su}.

Any differential operator commuting with $H$ is
called a quantum integral of the system.
One says that two quantum integrals are in involution if they commute
as differential operators.

The system defined by $H$ is called completely integrable if it has
$N$ quantum integrals in involution, $L_1,...,L_{N}$, which are
algebraically independent: every polynomial $P\in\Bbb
R[z_1,...,z_{N}]$ such that $P(L_1,...,L_{N})=0$
is identically zero.

 For $N>2$, the system defined by $H$ is not always
completely integrable.
However, for special choices of $U(x)$ it is known that it
is completely integrable for any $K$. For example, this is proved for
$U(x)=x^{-2}$ (rational case), $U(x)=\text{sinh}(x)^{-2}$
(trigonometric case), and
$U(x)=\wp(x|\tau)$, where $\wp$ is the Weierstrass elliptic function:
$$
\wp(x|\tau)=\frac{1}{x^2}+\sum_{(m,n)\in\Bbb Z^2\backslash\{(0,0)\}}
\biggl[\frac{1}{(x-m-n\tau)^2}-\frac{1}{(m+n\tau)^2}\biggr]\tag 2
$$
(elliptic case) (see \cite{OP} and references therein).
These results come from an explicit construction of quantum integrals
via the Lax matrix.

In this paper we propose a new construction
of the quantum integrals for $H$ in the trigonometric and elliptic cases
using representation theory of Lie algebras, and give a new
proof of the complete integrability theorem based on this construction.
In the trigonometric case, we construct a homomorphism
from the center of the universal enveloping algebra of $\glN$
to the algebra of differential operators in $N$ variables. We show
that the image of the second-order Casimir under this homomorphism is a
multiple of the Sutherland operator, which implies that
the images of higher Casimirs are its quantum integrals and thus
proves the integrability. In the elliptic case, we apply a similar
method to the center of the (completed) universal enveloping algebra
of the affine $\glN$ at the critical level $k=-N$, and thus produce
quantum integrals of the Hamiltonian (1) with the elliptic potential
$U=K\wp$.

If $U(x)$ is one of the above three choices, we call $H$ a
Calogero-Sutherland (CS) operator. If $H$ is a CS operator then
for any set of complex numbers $\Lambda_i$
the system of differential equations
$$
L_i\psi=\Lambda_i\psi,\ 1\le i\le N \tag 3
$$
is consistent and has $N!$ linearly independent solutions.
This system is called the eigenvalue problem for $H$.
The solutions of (3) for the rational and trigonometric cases
(unlike the elliptic case) have been studied in many papers
\cite{*} and are rather well understood.
 We will give a new construction of these
eigenfunctions for the trigonometric case as normalized traces of
intertwiners between representations
of $\frak{gl}_N$.

\centerline{\bf Acknowledgements} I would like to thank
my advisor Professor Igor Frenkel for systematic
inspiring discussions and I.Cherednik, D.Kazhdan, A.Kirillov Jr.,
F. Malikov, K.Styrkas, and A.Varchenko for many useful remarks and
suggestions.

\vskip .1in
\heading{\bf 1. Diagonalization of the Sutherland operator}\endheading
\vskip .1in

The Sutherland operator is the following differential operator:
$$
H=\sum_{j=1}^N\frac{\d^2}{\d x_j^2}+\sum_{1\le i<j\le
N}\frac{K}{\text{sinh}^2(x_i-x_j)},\ K\in\Bbb C
\tag 1.1
$$
The quantum integrals for this operator and their eigenfunctions
(zonal spherical functions) are known \cite{OP,HO}.
In this section we will give a new description
of the quantum integrals and eigenfunctions
 using representatin theory of $\glN$.

Let $\glN$ denote the Lie algebra of complex $N\times N$ matrices.
If $A\in\glN$, we will denote by $A_{ij}$ the entry at the
intersection of the $i$-th row and $j$-th column of $A$.
Also, $E_{ij}$
will denote the elementary matrices:
$(E_{ij})_{mn}=\delta_{im}\delta_{jn}$. For brevity we denote $E_{ii}$
by $h_i$.

Let $\l=(\l _1,...,\l _N)\in\Bbb C^N$, and let $\Ml$
be the Verma module over $\glN$ with highest weight $\lambda$, i.e.
the module generated by a highest weight vector $v_{\l}$
satisfying the defining relation
$$
Av_{\l}=\sum_{j=1}^N\lambda_jA_{jj}v_{\l},\tag 1.2
$$
whenever $A$ is an upper triangular matrix.
Let $\Ml^*$ denote the restricted dual module
to $\Ml$ -- the direct sum of dual spaces to the weight subspaces in
$\Ml$ with the action of $\glN$ defined by duality.

Let $\mu\in\Bbb C$. Define a module $\Vm$ over $\glN$ as follows.
 As a vector space, $\Vm$ is the space of
functions of the form $f(\bold
x)=(\prod_{j=1}^m\xi_j)^{\mu}p(\frac{\xi_1}{\xi_2},...,\frac{\xi_{N-1}}
{\xi_N})$,
where $p\in\Bbb C[y_1^{\pm 1},\dots,y_{N-1}^{\pm 1}]$ is a Laurent
polynomial, and the action $\phi$ of $\glN$ in $\Vm$ is
described
by the formula $\phi(E_{ij})=\xi_i\frac{\d}{\d \xi_j}-\mu\delta_{ij}$.

It is obvious that the set of weights of $\Vm$ is
$\{(\l _1,...,\l _N)\in\Bbb
Z^N:\sum\l _j=0\}$, and
every weight occurs with multiplicity 1.

Consider the completed tensor product $\Ml\hat\otimes \Vm=
\text{Hom}_{\Bbb C}(\Ml^*,\Vm)$. It has a natural structure of a
$\glN$-module.

\proclaim{Proposition 1.1} If $\Ml$ is irreducible then
there exists a unique up to a factor nonzero intertwining operator
$\Phi_{\lambda}:\Ml\to\Ml\hat\otimes \Vm$.
\endproclaim

\demo{Proof} We need to prove that the module
$\text{Hom}_{\Bbb C}(\Ml^*,\Vm)$ contains a singular vector $w_{\l}$
of weight $\lambda$ (a vector satisfying relation (1.2)),
and that such a vector is unique. Since $w_{\l}$ is invariant under
the Lie subalgebra ${\frak n}^+$ of strictly upper triangular matrices, it is
uniquely determined by its value on the lowest weight vector
$v_{\lambda}^*$ of the module $\Ml^*$. Because $\Ml$ is irreducible,
$\Ml^*=U({\frak n}^+)v_{\l}^*$, and therefore $w_{\l}(v_{\l}^*)$
can be any zero weight vector in $\Vm$. Since the zero weight subspace
of $\Vm$ is one dimensional, the vector $w_{\l}$ exists and is
unique up to a factor, Q.E.D.
$\blacksquare$\enddemo

We fix the normalization of the operator $\Phi_{\l}$ under which
$<v_{\lambda}^*,\Phi v_{\lambda}>=\prod \xi_j^{\mu}\in \Vm$.

Let $\rho=(\rho_1,...,\rho_N),\ \rho_j=\frac{N+1}{2}-j$.
Consider the function of $N$ variables
$$
\Psi_{\lambda}(z_1,...,z_N)=
\frac{\text{Tr}\mid_{\Ml}(\Phi_{\lambda}z_1^{h_1}\dots
z_N^{h_N})}{\text{Tr}\mid_{M_{-\rho}}(z_1^{h_1}\dots
z_N^{h_N})}.\tag 1.3
$$
This function takes values in $\Vm$ and has weight zero,
hence
$$
\Psi_{\l}=\psi_{\l}\prod \xi_j^{\mu},\tag 1.4
$$
 where $\psi_{\l}$ is a complex-valued function.

Let us introduce new coordinates $x_i$ such that $z_i=e^{2x_i}$,
and regard the function $\psi_{\l}$ as a function of $(x_1,...,x_N)$

If $a,b\in\Bbb C^N$, let $<a,b>=\sum_{j=1}^Na_jb_j$, and $a^2=<a,a>$.
Also, for brevity we use the notation $Z=z_1^{h_1}\dots z_N^{h_N}$,
and write $\text{Tr}_{\l}$ for $\text{Tr}\mid_{M_{\l}}$.

\proclaim{Theorem 1.2} If $K$ in formula (1.1) equals $-2\mu(\mu+1)$ then
the function $\psi_{\l}$ is an eigenfunction of the differential
operator $H$ defined by (1.1) with the eigenvalue $4(\l+\rho)^2$.
\endproclaim

\demo{Proof}
The Casimir element
$$
C_2=\sum_{i,j=1}^NE_{ij}E_{ji}\tag 1.5
$$
acts in $\Ml$ by multiplication by
$<\lambda,\lambda+2\rho>$. Therefore,
$$
\frac{\text{Tr}_{\l}(\Phi_{\lambda}C_2Z
)}{\text{Tr}_{{-\rho}}(Z
)}=<\l,\l+2\rho>\Psi_{\l}(z_1,...,z_N).\tag 1.6
$$
On the other hand, the left hand side of (1.6) can be written
in the form $\sum_{i,j=1}^N\Psi^{ij}_{\l}$, where
$$
 \Psi^{ij}_{\lambda}=
\frac{\text{Tr}_{\l}(\Phi_{\lambda}E_{ij}E_{ji}Z
)}{\text{Tr}_{{-\rho}}(Z
)}.\tag 1.7
$$
Let us express the terms $\Psi^{ij}_{\l}$ in terms of
$\Psi_{\l}$. First of all,
consider the function $F(z_1,...,z_N)=\text{Tr}_{{-\rho}}(Z
)$. This function is the character
of the module $M_{-\rho}$, therefore
$$
\gather
F(z_1,...,z_N)=\prod_{j=1}^Nz_j^{-\rho_j}\prod_{i>j}\biggl(1-\frac
{z_i}{z_j}\biggr)^{-1}=
\frac{(\prod_{j=1}^Nz_j)^{\frac{N-1}{2}}}
{\prod_{i>j}(z_i-z_j)},\\
\frac{z_i}{F}\frac{\d F}{\d z_i}=
\biggl(\frac{N-1}{2}-\sum_{j\ne
i}\frac{z_i}{z_i-z_j}\biggr).
\tag 1.8\endgather
$$

We consider two cases.

1) $i=j$. Then, using (1.8), we obtain
$$
\gather
\Psi^{ii}_{\l}=\frac{1}{F} \biggl(z_i\frac{\d}{\d
z_i}\biggr)^2(F\Psi_{\l})=\\
\biggl(z_i\frac{\d}{\d
z_i}\biggr)^2\Psi_{\l}+\biggl((N-1)z_i-\sum_{j\ne
i}\frac{2z_i^2}{z_i-z_j}\biggr)\frac{\d \Psi_{\l}}{\d
z_i}+\frac{\biggl(z_i\frac{\d}{\d z_i}\biggr)^2F}{F}\Psi_{\l}.\tag 1.9
\endgather
$$

2) $i\ne j$. Then we have
$$
\gather
\Psi^{ij}_{\l}=F^{-1}\text{Tr}_{\l}
(\Phi_{\lambda}E_{ij}E_{ji}Z
)=\\
F^{-1}\text{Tr}_{\l}(E_{ij}\Phi_{\lambda}E_{ji}Z
)+
F^{-1}E_{ij}\text{Tr}_{\l}(\Phi_{\lambda}E_{ji}Z
)=\\
F^{-1}\text{Tr}_{\l}(\Phi_{\lambda}E_{ji}Z
E_{ij})+
F^{-1}E_{ij}\text{Tr}_{\l}(\Phi_{\lambda}E_{ji}Z
)=\\
\frac{z_i}{z_j}
F^{-1}\text{Tr}_{\l}(\Phi_{\lambda}E_{ji}E_{ij}Z
)+
F^{-1}E_{ij}\text{Tr}_{\l}(\Phi_{\lambda}E_{ji}Z
)=\\
\frac{z_i}{z_j}
F^{-1}\text{Tr}_{\l}(\Phi_{\lambda}E_{ij}E_{ji}Z
)+
\frac{z_i}{z_j}
F^{-1}\text{Tr}_{\l}(\Phi_{\lambda}(E_{jj}-E_{ii})Z
)\\
+F^{-1}E_{ij}\text{Tr}_{\l}(\Phi_{\lambda}E_{ji}Z
)=\\
\frac{z_i}{z_j}\Psi^{ij}_{\l}+
F^{-1}\frac{z_i}{z_j}\biggl(z_j\frac{\d}{\d z_j}-z_i\frac{\d}{\d
z_i}\biggr)
(F\Psi_{\l})+
F^{-1}E_{ij}\text{Tr}_{\l}(\Phi_{\lambda}E_{ji}Z
).\tag 1.10
\endgather
$$
Formula (1.10) is a linear equation on $\Psi^{ij}_{\l}$. Solving this
equation, we obtain
$$
\Psi^{ij}_{\l}=F^{-1}\frac{z_i}{z_j-z_i}
\biggl(z_j\frac{\d}{\d z_j}-z_i\frac{\d}{\d z_i}\biggr)
(F\Psi_{\l})+\frac{F^{-1}z_j}{z_j-z_i}E_{ij}
\text{Tr}_{\l}(\Phi_{\lambda}E_{ji}Z).\tag 1.11
$$
It remains to compute
$\text{Tr}_{\l}(\Phi_{\lambda}E_{ji}Z)$. We have
$$
\gather
\text{Tr}_{\l}(\Phi_{\lambda}E_{ji}Z)=
\text{Tr}_{\l}(E_{ji}\Phi_{\lambda}Z)+
E_{ji}\text{Tr}_{\l}(\Phi_{\lambda}Z)=
\text{Tr}_{\l}(\Phi_{\lambda}ZE_{ji})+FE_{ji}\Psi_{\l}=\\
\frac{z_j}{z_i}\text{Tr}_{\l}(\Phi_{\lambda}E_{ji}Z)+FE_{ji}\Psi_{\l},
\tag 1.12\endgather
$$
which implies
$$
\text{Tr}_{\l}(\Phi_{\lambda}E_{ji}Z)=\frac{z_i}{z_i-z_j}FE_{ji}\Psi_{\l}.\tag
1.13
$$
Combining (1.11) and (1.13), we deduce
$$
\Psi^{ij}_{\l}=F^{-1}\frac{z_i}{z_j-z_i}
\biggl(z_j\frac{\d}{\d z_j}-z_i\frac{\d}{\d z_i}\biggr)
(F\Psi_{\l})-\frac{z_iz_j}{(z_i-z_j)^2}E_{ij}E_{ji}\Psi_{\l}
.\tag 1.14
$$
It is easy to see that the operator $E_{ij}E_{ji}$ acts in the zero
weight subspace of $\Vm$ by multiplication by $\mu(\mu+1)$. Therefore,
(1.14) can be rewritten as follows:
$$
\gather
\Psi^{ij}_{\l}=\frac{z_i}{z_j-z_i}
\biggl(z_j\frac{\d}{\d z_j}-z_i\frac{\d}{\d z_i}\biggr)
\Psi_{\l}\\
+\biggl[\frac{z_i}{z_j-z_i}
\biggl(z_j\frac{\d}{\d z_j}-z_i\frac{\d}{\d z_i}\biggr)\log F
-\mu(\mu+1)\frac{z_iz_j}{(z_i-z_j)^2}\biggr]\Psi_{\l}
.\tag 1.15\endgather
$$
Now, summing up equations (1.9) for all $i$ and equations (1.14)
for all $i\ne j$, and using (1.8) and the identity
$$
\sum_{i\ne j}\frac{z_i}{z_j-z_i}\biggl(z_j\frac{\d}{\d
z_j}-z_i\frac{\d}{\d z_i}\biggr)=
\sum_{i\ne
j}\biggl(\frac{2z_i^2}{z_i-z_j}-(N-1)z_i\biggr)\frac{\d}{\d z_i},\tag 1.16
$$
we obtain
$$
\gather
\frac{\text{Tr}_{\l}(\Phi_{\lambda}C_2Z
)}{\text{Tr}_{{-\rho}}(Z
)}=\sum_{i=1}^N
\biggl(z_i\frac{\d}{\d
z_i}\biggr)^2\Psi_{\l}+\\
\biggl[\sum_{i=1}^N\frac{\biggl(z_i\frac{\d}{\d z_i}\biggr)^2F}{F}-
2\sum_{i=1}^N\biggl(\frac{z_i\frac{\d}{\d z_i}F}{F}\biggr)^2
-2\sum_{i<j}\mu(\mu+1)\frac{z_iz_j}{(z_i-z_j)^2}\biggr]\Psi_{\l}.\tag 1.17
\endgather
$$

{}From formula (1.8) it follows that
$$
F^{-1}(z_1,...,z_N)=\text{det}(z_i^{-\rho_j}).\tag 1.18
$$
Using this identity, we find
$$
\sum_{i=1}^N\biggl(z_i\frac{\d}{\d
z_i}\biggr)^2F^{-1}=<\rho,\rho>F^{-1}.\tag 1.19
$$
Substituting (1.19) into (1.17), we get
$$
\gather
\frac{\text{Tr}_{\l}(\Phi_{\lambda}C_2Z
)}{\text{Tr}_{{-\rho}}(Z
)}=\sum_{i=1}^N
\biggl(z_i\frac{\d}{\d
z_i}\biggr)^2\Psi_{\l}-
<\rho,\rho>\Psi_{\l}
-2\sum_{i<j}
\mu(\mu+1)\frac{1}{\frac{z_i}{z_j}+\frac{z_j}{z_i}-2}\Psi_{\l}.\tag
1.20
\endgather
$$
Rewriting equation (1.20) in the new variables $x_i$ such that
$z_i=e^{2x_i}$,
we obtain
$$
\gather
\frac{\text{Tr}_{\l}(\Phi_{\lambda}C_2Z
)}{\text{Tr}_{{-\rho}}(Z
)}=\frac{1}{4}\sum_{i=1}^N
\frac{\d^2 \Psi_{\l}}{\d
x_i^2}-
<\rho,\rho>\Psi_{\l}
-\frac{\mu(\mu+1)}{2}\sum_{i<j}\frac{1}{\text{sinh}^2(x_i-x_j)}\Psi_{\l}.\tag
1.21
\endgather
$$
Comparing (1.6) and (1.21), we finally get
$$
\sum_{i=1}^N
\frac{\d^2 \psi_{\l}}{\d
x_i^2}-2\mu(\mu+1)\sum_{i<j}\frac{1}{\text{sinh}^2(x_i-x_j)}\psi_{\l}=
4(\l+\rho)^2\psi_{\l}.\tag 1.22
$$
The theorem is proved.
$\blacksquare$\enddemo

\vskip .1in
\centerline{\bf 2. Quantum integrals of the Sutherland
operator}
\vskip .1in

Let us now find the quantum integrals of the Sutherland operator.
For this purpose we will use the higher Casimir elements of the Lie
algebra $\glN$.

Let $Y$ be any element of the universal enveloping algebra
$U(\glN)$.
Then we can consider the following $\Vm$-valued
function:
$$
\Psi_{\lambda}(Y|z_1,...,z_N)=
\frac{\text{Tr}\mid_{\Ml}(\Phi_{\lambda}Yz_1^{h_1}\dots
z_N^{h_N})}{\text{Tr}\mid_{M_{-\rho}}(z_1^{h_1}\dots
z_N^{h_N})}.\tag 2.1
$$

\proclaim{Proposition 2.1}
There exists a differential operator $\Cal L_Y$ in $z_1,...,z_N$
whose coefficients are $U(\glN)$-valued functions, such that
$$
\Psi_{\l}(Y|z_1,...,z_N)=\Cal L_Y\Psi_{\l}(z_1,...,z_N).\tag 2.2
$$
\endproclaim

\demo{Proof} We say that $Y$ is of order $\le m$ if it is a sum of
monomials including $\le m$ factors from $\glN$. The proof is by
induction in the order of $Y$. For order $0$ the statement is
obvious. If $Y$ is of a positive order $m$ then we may assume
that $Y$ is a monomial of the form $Y=XE_{ij}$. Since
$X$ is a monomial of a lower order than $Y$, we may assume that
the operator $\Cal L_X$ is already defined.

Again, we consider two cases. If $i=j$ then
$$
\Psi_{\l}(Y|z_1,...,z_N)=F^{-1}z_i\frac{\d}{\d z_i}
(F\Psi_{\l}(X|z_1,...,z_N)),\tag 2.3
$$
so we can set $\Cal L_Y=F^{-1}\circ z_i\frac{\d}{\d z_i}
\circ F\circ \Cal L_X$.

If $i\ne j$ then
let $U=[X,E_{ij}]$.
Clearly, the order of $U$
is $\le m-1$, so we can assume that $\Cal L_U$ has been defined.
Then we have
$$
\gather
\Psi_{\l}(Y)=\Psi_{\l}(E_{ij}X)+\Psi_{\l}(U)=
E_{ij}\Psi_{\l}(X)+\frac{z_i}{z_j}\Psi_{\l}(Y)+\Cal L_U\Psi_{\l}=\\
(E_{ij}\Cal L_X+\Cal L_U)\Psi_{\l}+\frac{z_i}{z_j}\Psi_{\l}(Y),\tag 2.4
\endgather
$$
which implies
$$
\Psi_{\l}(Y)=\frac{z_j}{z_j-z_i}(E_{ij}\Cal L_X+\Cal L_U)\Psi_{\l}.\tag 2.5
$$
This proves the step of induction Q.E.D.
$\blacksquare$\enddemo

Let $Y\in U(\glN)$ be
of weight zero, i.e. such that $[h_i,Y]=0$ for all $i$.
Then $\Psi_{\l}(Y|z_1,...,z_N)=\psi_{\l}(Y|x_1,...,x_N)
\prod_{i=1}^N\xi_i^{\mu}$, where $\psi$ is a complex-valued function.

\proclaim{Proposition 2.2} If $Y$ is of weight zero
then there exists a unique scalar-valued differential
operator $D_Y$ in the variables $x_1,...,x_N$ whose coefficients
are rational functions of $e^{2x_i},\ 1\le i\le N$ such that
$$
\psi_{\l}(Y|x_1,...,x_N)=D_Y\psi_{\l}(x_1,...,x_N)\tag 2.6
$$
for a generic $\lambda$.
\endproclaim

\demo{Proof} If $Y$ is of weight 0 then $\Cal L_Y$ preserves weight, so
its coefficients map the zero weight subspace in $\Vm$ to itself.
Since the zero weight
subspace is one-dimensional, the coefficients of $\Cal L_Y$
become  scalars when restricted to this space.
In this way the operator $\Cal L_Y$ turns into
a scalar differential operator $D_Y$
which satisfies (2.6).

To establish the uniqueness, it is enough to prove the following:

{\bf Lemma 2.3. }
Any differential operator $D$  annihilating $\psi_{\l}$ for a
generic $\l$ whose coefficients
are rational functions in $z_i=e^{2x_i},\ 1\le i\le N$
must be identically equal to $0$.

{\it Proof. }
Since the functions $\psi_{\l}$ are homogeneous in $z_1,...,z_N$,
any homogeneous component of $D$ also annihilates them. Therefore,
we may assume that $D$ is homogeneous. We can also assume that $D$ is
 of degree $0$ (otherwise we can multiply it by a suitable power of
$z_1$)
and that its coefficients
are polynomials in $y_i=z_i/z_{i+1}$ (otherwise we can multiply $D$ by the
common denominator of its coefficients).
Let us write $D$ in the form
$$
D=\sum_{r_j\ge 0} y_1^{r_1}\dots y_{N-1}^{r_{N-1}}
D_{r_1,...,r_{N-1}},\tag 2.7
$$
 where $D_{r_1,...,r_{N-1}}=p_{r_1,...,r_{N-1}}(\frac{\d}{d
x_1},...,\frac{\d}
{\d x_N}$, where $p_{r_1,...,r_{N-1}}$ are polynomials.
Let $k$ be the smallest integer for which there
exist $r_1,...,r_{N-1}$  with $\sum r_j=k$ such that
$D_{r_1,...,r_{N-1}}\ne 0$. Since
$\psi_{l}\in \prod e^{2x_i(\lambda_j+\rho_j)}\Bbb C[[y_1,...,y_{N-1}]]$,
for this set of $r_j$ we have
$$
D_{r_1,...,r_{N-1}} e^{<2\bold x,\lambda+\rho>}=0,\ \bold
x=(x_1,...,x_N).\tag 2.8
$$
This implies that $p_{r_1,...,r_{N-1}}(2(\l+\rho))=0$ for a generic
$\l$, which means that $p_{r_1,...,r_{N-1}}$ is identically zero
 -- a contradiction.
This completes the proof of the lemma and the proposition.
$\blacksquare$\enddemo

It is well known that the center of
$U(\glN)$ is freely generated by the Casimir elements
$$
C_m=\sum_{j_1,...,j_m=1}^NE_{j_1j_2}\dots E_{j_{m-1}j_m}E_{j_mj_1},\
1\le m\le N.\tag 2.9
$$
Define the differential operators
$$
L_j=D_{C_j},\ 1\le j\le N\tag 2.10
$$

\proclaim{Proposition 2.4}

(i) $L_1=\frac{1}{2}
\sum_{j=1}^N\frac{\d}{\d x_j}$, $L_2=\frac{H}{4}+<\rho,\rho>$.

(ii) For any $1\le j\le N$ the function
$\psi_{\l}$ is an eigenfunction of $L_j$ with
an eigenvalue  $p_j(\l+\rho)$, where $p_j$ is a symmetric polynomial
of degree $j$.

(iii) $[L_i,L_j]=0$ for all $i,j$.

(iv) The symbol of $L_m$ is $2^{-m}\sum_{i=1}^N\frac{\d^m}{\d x_i^m}$.

(v) The operators $L_j$ and the polynomials $p_j$ are algebraically
independent.

(vi) The operators $L_i$ are invariant under permutations
of $x_1,...,x_N$.
\endproclaim

\demo{Proof}
(i) The formula for $L_1$ is obvious. The formula for $L_2$
follows from Theorem 1.2.

(ii) The element $C_i$ acts in $M_{\lambda}$ by multiplication by a
scalar polynomially dependent on $\l$.
This scalar is a symmetric polynomial in $\l+\rho$, since
the action of $C_i$ is the same in $\Ml$ and
$M_{\sigma(\l+\rho)-\rho}$
for any permutation $\sigma$ (proof: the first module contains the
second one when $\l$ is a dominant integral weight, and any two
polynomials coinciding on dominant integral weights coincide
identically). Denote this symmetric polynomial by $p_i$. Then
(2.10) implies that $L_i\psi_{\l}=p_i(\l+\rho)\psi_{\l}$.

(iii) Statement (ii) implies that $[L_i,L_j]\psi_{\l}=0$ for a
generic $\l$. According to Lemma 2.3, this means that $[L_i,L_j]=0$.

(iv) It is easy to show that
$$
C_m=\sum_{j=1}^N h_j^m+\sum_{i<j}X_{ij}E_{ij}+Y,\tag 2.11
$$
where $X_{ij}, Y$ are elements of $U(\glN)$ of order $\le m-1$.
This fact combined with the proof of Proposition 2.1 shows
that the symbol of $L_m$ will be the same as the symbol of
$D_{\sum_{j=1}^N h_j^m}$ (since the rest of the terms in (2.11) will give
lower order contributions), i.e. it will equal
$2^{-m}\sum_{i=1}^N\frac{\d^m}{\d x_i^m}$.

(v) Since the symbols of $L_j$ are algebraically independent
(they are elementary symmetric functions), so are $L_j$ themselves.
Therefore, so are $p_j$: if there existed a nonzero polynomial $Q$
such that $Q(p_1,...,p_N)$ is identically zero, then
$Q(L_1,...,L_N)$ would annihilate $\psi_{\l}$ for a generic $\l$,
which, by Lemma 2.3, would mean that this operator is zero, so that
from the algebraic independence of $L_j$ one gets $Q=0$.

(vi) Let $\sigma$ be any transposition. Then for any $Y$ one has
$\Cal L_{\sigma Y\sigma^{-1}}=\sigma \Cal L_Y^{\sigma}\sigma^{-1}$,
where $\Cal L_Y^{\sigma}$ is defined by the formula:
$$
(\Cal L^{\sigma}f) (x_{\sigma(1)},...,x_{\sigma(N)})=\Cal L
f(x_{\sigma(1)},...,x_{\sigma(N)}).\tag 2.12
$$
Therefore, $D_{\sigma Y\sigma^{-1}}=D_Y^{\sigma}$.
On the other hand, $\sigma C_m\sigma^{-1}=C_m$, which implies
that $L_m^{\sigma}=L_m$.
$\blacksquare$\enddemo

We can now consider the eigenvalue problem (3) associated with the
collection of operators $\{L_j\}$. Recall that this problem is to find
a basis of the $N!$-dimensional space of solutions of the system
of differential equations $L_i\psi=\Lambda_i\psi$, $1\le i\le N$,
where $\Lambda_i$ are some given complex numbers.

\proclaim{Proposition 2.5} Let $\lambda$ be a solution of
the system of algebraic equations
$$
p_i(\lambda+\rho)=\Lambda_i,\ 1\le i\le N.\tag 2.13
$$
Suppose that $\lambda_i+\rho_i-\lambda_j-\rho_j$ is not an
integer for any $i\ne j$. Then the functions
$\{\psi_{\sigma(\l+\rho)-\rho},\ \sigma\in S_N\}$,
(where $S_N$ is the symmetric group) form a basis in the space of
solutions of system (3).
\endproclaim

\demo{Proof} First of all, all the functions
$\{\psi_{\sigma(\l+\rho)-\rho}\}$ are defined since
all the modules $M_{\sigma(\l+\rho)-\rho}$
are irreducible. Next, the weights ${\sigma(\l+\rho)-\rho}$
are different for different $\sigma$, hence the functions
$\{\psi_{\sigma(\l+\rho)-\rho}\}$ are linearly independent
(this follows from the asymptotics $\psi_{\l}\sim \prod
z_j^{\lambda_j+\rho_j}$, $z_i/z_{i+1}\to 0, 1\le i\le N-1$).
And finally, all the functions
$\{\psi_{\sigma(\l+\rho)-\rho}\}$ are solutions of (3):
$\psi_{\l}$ is a solution of (3) by the definition of $\l$, and the
rest are solutions because the operators $L_i$ and the
polynomials $p_i$ are symmetric
and therefore $\lambda^{\sigma}=
{\sigma(\l+\rho)-\rho}$ satisfies (2.13) for any
$\sigma\in S_N$.
$\blacksquare$\enddemo

{\bf Remark.} In 1976 I.Frenkel computed the Laplace's operator
in the space of functions on $(SU(N)\times SU(N))/SU(N)_{diag}$ with
values in a symmetric power of the fundamental
representation of $SU(N)$ equivariant with respect to
the left diagonal action of $SU(N)$, and found that it was equal to
the Sutherland operator (with sin instead of sinh) and $K=-2m(m+1)$,
where $m$ is an integer (I.Frenkel, unpublished work). Proposition 2.5
can be regarded as a generalization of this result.

\vskip .1in
\heading
{\bf 3. Affine Lie algebras, vertex operators, correlation functions,
and regular expressions}
\endheading
\vskip .1in

Recall the definition of affine Lie algebras.

The affine Lie algebra $\widehat{\frak{gl}_N}$ is defined as follows:
as a vector space \linebreak
$\widehat{\frak{gl}_N}=
\glN\otimes \Bbb C[t,t^{-1}]\oplus\Bbb Cc$,
and the commutator is given by
$$
[A(t)+\alpha c,B(t)+\beta c]=[A(t),B(t)]+\text{Res}_{t=0}
(t^{-1}\text{Tr}(A^{\prime}(t)B(t))),\tag 3.1
$$
where
$A(t)$, $B(t)$ are matrix-valued
trigonometric polynomials.

The principal gradation on $\widehat{\frak{gl}_N}$ assigns degree
$j-i+Nm$ to the element $E_{ij}t^m$, and degree zero to $c$.
The degree of an element $a$ is denoted by $\deg(a)$.

Define the Lie subalgebra $\g$ in $\widehat{\frak{gl}_N}$
to be the Lie algebra of all elements
$A(t)+\alpha c\in\widehat{\frak{gl}_N}$ such that
$\text{Tr}(A(t))$
is a constant (does not depend on $t$). This algebra is isomorphic to
the direct sum of the affine Lie algebra $\hatslN$ and the one-dimensional
abelian Lie algebra.
It inherits the grading from $\widehat{\frak{gl}_N}$.
We denote by $\hatglN^{\pm}$ the subalgebras in $\hatglN$ spanned by
the elements of positive (respectively negative) degree.

We will work with two kinds of modules over $\hatglN$ -- Verma modules and
evaluation modules.

Let $\lambda\in\Bbb C^N$ and $V$ be a $\g$-module. We say that a
nonzero vector $v\in V$ is of weight $\lambda$ if $h_iv=\lambda_iv$.
We say that a nonzero element
$a\in U(\g)$ is of weight $\l$ if $[h_i,a]=\lambda_ia$.

Let $\lambda\in\Bbb C^N$ and $k\in \Bbb C$. The Verma module
$M_{\lambda,k}$ over $\hatglN$ is defined by a single generator
$\vlk$ and the relations:
$$
(A(t)+\alpha c)\vlk=(\sum_{j=1}^NA_{jj}(0)\lambda_j+\alpha k)\vlk, \
A(t)+\alpha c\in\g\tag 3.2
$$
if $A(t)$ is regular at $t=0$, and $A(0)$ is upper triangular.
This module is a highest weight module which
is irreducible for generic $\lambda$ and $k$.

The principal gradation operator $\d$ on
$M_{\lambda,k}$ is defined by the relations:
$$
\d\vlk=0,\ [\d, a]=na\text{ if }a\in\g,\ \deg(a)=n.\tag 3.3
$$
A vector $w\in\Mlk$ is called homogeneous if it is an eigenvector of $\d$.
We say that a vector $w\in\Mlk$ is of degree $n$ ($\text{deg}(w)=n$)
if $\d w=nw$.

It is easy to check that
the only vectors of degree $0$ in $\Mlk$ are multiples of $\vlk$, and
the
rest of
homogeneous vectors have negative degrees. Also, the space
of vectors of a fixed degree $n$ is finite dimensional.
This space is called the homogeneous subspace of degree $n$,
and denoted by $\Mlk[n]$.

We will need to work with the completion $\hat\Mlk$
of the Verma module $\Mlk$
with respect to the principal gradation.
$\hat\Mlk$ is defined as the space of infinite sums of the form
$\sum_{n\le 0}w_n$, where $w_n$ is a vector of degree $n$.
Clearly, the action of $\hatglN$ in $\Mlk$ naturally extends to
$\hat\Mlk$.

Define also the restricted dual module $\Mlk^{\prime}$ to $\Mlk$ --
the direct sum of the dual spaces to the homogeneous subspaces in
$\Mlk$, with the action of $\hatglN$ defined by duality.
This module is a lowest weight module. Denote its lowest weight vector
by $\vlk^{\prime}$.
It is obvious that $\hat\Mlk$ is the (usual) dual module
to $\Mlk^{\prime}$.

Now let us define evaluation modules. Let $\mu\in\Bbb C$. Define the
module $V_{\mu}$ over $\hatglN$ to be the pullback of the module
$\Vm$ over $\glN$ defined in Section 1 with respect to the evaluation
map $\pi: \g\to\glN$ given by $\pi(A(t)+\alpha c)=A(1)$

Now let us introduce vertex operators. Throughout this paper by
a vertex operator we mean a $\hatglN$- intertwining operator
$\Phi:\Mlk\to\Mlk\hat\otimes\Vm$, where
$\Mlk\hat\otimes\Vm$ denotes the space of formal sums of the form
$\sum_{j=0}^{\infty}u_j\otimes v_j$, $u_j\in\Mlk[-j]$, $v_j\in\Vm$.

\proclaim{Proposition 3.1} If $\Mlk$ is irreducible then there exists
a unique nonzero vertex operator up to a scalar factor.
\endproclaim

\demo{Proof} In order to construct $\Phi$, we first need to construct
the vector $w=\Phi \vlk\in \Mlk\hat\otimes \Vm$.
This vector has to be annihilated by the subalgebra $\hatglN^+$
and satisfy the condition $h_iw=\l_iw$. To construct such a vector
is the same as to construct a map $w:\Mlk^{\prime}\to \Vm$ which commutes
with the action of $\hatglN^+$ and maps $\vlk^{\prime}$ to a zero
weight vector. Since $\Mlk$ is irreducible, $\Mlk^{\prime}$ is freely
generated by $\vlk^{\prime}$ over $U(\hatglN^+)$, which implies that
the map $w$ is uniquely determined once we know the image of
$\vlk^{\prime}$ under this map. Because the space of zero weight
vectors in $\Vm$ is one-dimensional, $w$ is defined uniquely
up to a factor.

Finally, since $\Phi$ is an intertwiner, and $\Mlk$ is freely
generated by $\vlk$ over $U(\hatglN^-)$, $\Phi$ is uniquely determined
once we know the image of $\vlk$ under it, i.e. once we know $w$.
Therefore, $\Phi$ is unique up to a factor.
$\blacksquare$\enddemo

Let $A:\Mlk\to\hat\Mlk$ be any linear operator.
It is obvious that
we can uniquely represent the operator $A$ as an infinite sum
$$
A=\sum_{m\in\Bbb Z}A[m],\tag 3.4
$$
where $A[m]=\oplus_{j}A[m]_j$, and $A[m]_j: \Mlk[j]\to
\Mlk[j+m]$ are linear operators: for a homogeneous vector
$u$ of degree $j$ in $\Mlk$ the vector $A[m]_{j}u$ is defined as the degree
$j+m$ component of $A u$.

Let $z_j,\ 1\le j\le N$,
and $q$ be formal variables. Let $\bold z=(z_1,...,z_N)$.
Then we can consider the formal series
$$
F_{\lambda,k}(A,\bz,q)=\text{Tr}\mid_{\Mlk}(A
q^{-\d}\prod_{j=1}^Nz_j^{h_j})=\sum_{m\ge
0}q^m\text{Tr}\mid_{\Mlk[m]}
(A[0]_m
\prod_{j=1}^Nz_j^{h_j}).\tag 3.5
$$
Every coefficient of this formal series is a function
of the form $\prod_{j=1}^Nz_j^{\lambda_j}\cdot p(\frac{z_1}{z_2},...,
\frac{z_{N-1}}{z_N})$, where
$p$ is a Laurent polynomial.

Now let us define the (modified) expectation value
(or 1-point correlation function)
of an operator A as follows:
$$
<A>_{\l,k}(\bz)=
\frac{F_{\l,k}(A,\bz,q)}{
F_{-\rho,k}(\text{Id},
\bz,q)},\ \rho=(\rho_1,...,\rho_N),\ \rho_j=\frac{N+1}{2}-j
.\tag
3.6
$$
Then $<A>_{\l,k}\in\Bbb C[(z_1/z_2)^{\pm 1},...,(z_{N-1}/z_N)^{\pm 1}
][[q]]$ for any $A$.

Let $\theta:\Vm\to\Bbb C$ be a zero weight linear function. This
function is unique up to a factor.
Let the operator $\Phi_0:\Mlk\to\hat\Mlk$ be defined by the condition:
$\Phi_0u=(\text{Id}\otimes\theta)(\Phi u)$, $u\in\Mlk$.
The main object of our study will be the 1-point correlation function
of this operator:
$$
\pslk(\bz,q)=<\Phi_0>_{\lambda,k}.\tag 3.7
$$

Finally, let us define the algebra of regular expressions,
$\Uloc$.

 Let $X_{\pm}(j)$ be
an eigenbasis for $\{h_i\}$ and $\d$ in $\hatglN^{\pm}$,
respectively, such that the eigenvalue of $\d$ on $X_{\pm}(j)$
is a monotonic function of $j$. Let us call an expression of the form
$$
Y=X_-(m)^{n_m}\dots X_-(1)^{n_1}h_1^{p_1}\dots h_N^{p_N}X_+(1)^{r_1}
\dots X_+(s)^{r_s}c^M \in U(\hatglN)\tag 3.8
$$
a standard monomial.
Let us call the number
$$
\text{ord}(Y)=\sum_{j=1}^mn_j+\sum_{j=1}^Np_j+\sum_{j=1}^sr_j+M\tag 3.9
$$
the order of the standard monomial $Y$, and the number
$$
\text{sdeg}(Y)=\sum_{j=1}^sr_j\text{deg}(X_j^+)\tag 3.10
$$
the subsidiary degree of $Y$.
 According to the
Poincare-Birkhoff-Witt theorem, the
standard monomials form a basis in $U(\hatglN)$.

\proclaim{Definition} 1) A sum $\sum_{m=1}^rb_mY_m$, where
$b_m\in\Bbb C^*$ and $Y_m$ are distinct standard monomials, is called
a regular expression of order $\le n$ if
$\text{ord}(Y_m)\le n$ for all $m$.

2) A series $\sum_{m=1}^{\infty}b_mY_m$, where
$b_m\in\Bbb C^*$ and $Y_m$ are distinct standard monomials, is called
a regular expression of order $\le n$ if

(i) $\text{ord}(Y_m)\le n$ for all $m\ge 1$;

(ii) $\lim_{m\to\infty}\sdeg(Y_m)=\infty$.

3) A regular expression is of order $n$ if it is of order $\le n$ but
not of order $\le n-1$.
\endproclaim

It is easy to show that regular expressions form an associative
algebra. This algebra is a completion of
$U(\hatglN)$ and is denoted by $\Uloc$.

The main property of the algebra $\Uloc$ is that the action of
$U(\hatglN)$ in Verma modules over $\hatglN$ can be naturally
extended to $\Uloc$: if $R=\sum_{m=1}^{\infty}b_mY_m$ is a regular
expression then for any $w\in \Mlk$ $Rw=\sum_{m=1}^nb_mY_mw$
if $n$ is large enough.

\vskip .1in
\heading
{\bf 4. The mapping $\chi$}
\endheading

Let $\Cal {DO}_N$ be the algebra of differential operators in $z_1,...,z_N$
with Laurent polynomial coefficients (over $\Bbb C$). Let $DO_N=\Cal
{DO}_N[[q]]$. Let $\Cal A=\text{End}(\Vm)\otimes DO_N$.

We will consider expressions of the form
$<\Phi Y>_{\l,k}$, where $Y$ is a linear combination of standard
monomials. Such expressions are (formal) functions of $\bz,q$ with
values in $\Vm$. For the sake of brevity, we will write
$<A>$ instead of $<A>_{\l,k}$.

\proclaim{Theorem 4.1}
Let $Y$ be a finite linear combination of standard monomials of order $\le n$,
and let $k$ be a complex
number. Then:

1) There exists a
differential operator $L_Y(k)\in \Cal A$ such that
for all $\lambda$ for which $M_{\l,k}$ is irreducible
$$
<\Phi Y>=L_Y(k)<\Phi>,\tag 4.1
$$
satisfying the following conditions:

2) The operator $L_Y(k)$ is of order $\le n$ as a differential
operator;

3) The operator $L_Y(k)$ is a polynomial in $k$ of
degree $\le n$;

4) If $Y$ is a standard monomial and $\sdeg(Y)=s$ then
$L_Y=q^s\tilde L_Y$, where $\tilde L_Y\in \Cal A$.
\endproclaim

\demo{Proof} The proof of
of the theorem is by induction in the order of $Y$.
More precisely, we will construct $L_Y$ for
$Y$ of order $n$ from $L_J$ for standard monomials $J$ of lower
orders.

For order 0, $Y=1$, and we can take $L=1$. Assume that
$Y$ is a standard monomial of order $n>0$. Consider the last factor
$X\in\hatglN$
of this monomial, so that $Y=Y'X$, where $Y'$ is a standard monomial
of order $n-1$. Since $\ord(Y')=n-1$, we already
know that there exists an operator $L_{Y'}$ of order $\le n-1$
which is a polynomial in $k$ of degree $\le n-1$, such that $<\Phi
Y'>=L_{Y'}<\Phi>$.

There are three possibilities for $X$.

1. $X=c$. In this case we have $<\Phi Y>=k<\Phi Y'>$.
Then we can set $L_Y=kL_{Y'}$. This operator
satisfies (4.1). It also satisfies statements
2,3,4 of the theorem because so does $L_{Y'}$.

2. $X=X_{\pm}[j]$. In this case let $T=[Y'X]$.
It is clear that $T$ is a
finite linear combination of standard monomials  of order $\le n-1$,
so we may assume that the operator $L_T$ is already defined.
Then, using the intertwining property of $\Phi$ ($\Phi X=(X\otimes
1+1\otimes X)\Phi$), we can write
$$
<\Phi Y>=<\Phi Y'X>=<\Phi XY'>+<\Phi T>=<X\Phi Y'>+\pi(X)<\Phi Y'>+<\Phi
T>.\tag 4.2
$$
On the other hand, we have
$$
\gather
<X\Phi Y'>=\frac{\text{Tr}\mid_{\Mlk}(X\Phi Y'
q^{-\d}\prod_{j=1}^Nz_j^{h_j})}{F_{-\rho,k}}=\\
\frac{\text{Tr}\mid_{\Mlk}(\Phi Y'
q^{-\d}\prod_{j=1}^Nz_j^{h_j}X)}{F_{-\rho,k}}=
q^{-\deg(X)}\prod_{i=1}^Nz_i^{a_{i}}<\Phi
Y>.\tag 4.3
\endgather
$$
where $a_{i}$ are integers defined by $[h_i,X]=a_{i}X$.
Relations (4.2), (4.3) show that
$$
<\Phi Y>=
q^{-\deg(X)}\prod_{i=1}^Nz_i^{a_{i}}<\Phi
Y>+(\pi(X)L_{Y'}+L_T)<\Phi>,\tag 4.4
$$
which implies that
$$
<\Phi Y>=\frac{1}{1-q^{-\deg(X)}\prod_{i=1}^Nz_i^{a_{i}}}
(\pi(X)L_{Y'}+L_T)<\Phi>,\tag 4.5
$$
where $\frac{1}{1-q^mx}$ denotes $\sum_{p\ge 0}q^{mp}x^p$ if $m>0$,
and $-\sum_{p\ge 1}q^{-mp}x^{-p}$ if $m<0$. This means that
the differential operator
$$
L_Y=\frac{1}{1-q^{-\deg(X)}\prod_{i=1}^Nz_i^{a_{i}}}
(\pi(X)L_{Y'}+L_T)\tag 4.6
$$
satisfies (4.1). The fact that it satisfies statements 2,3,4 follows
from the validity of these statements for $L_{Y'}$ and $L_T$.

3. $X=h_i$. Then we have
$$
<\Phi Y>=\frac{z_i}{F_{-\rho,k}}\frac{\d}{\d z_i}(F_{-\rho,k} <\Phi Y'>),\tag
4.7
$$
which implies that the operator
$$
L_Y=\frac{z_i}{F_{-\rho,k}}\frac{\d}{\d z_i}\cdot
F_{-\rho,k}L_{Y'}\tag 4.8
$$
satisfies (4.1). The fact that it also satisfies statements 2,3,4,
as before, follows from the validity of these statements for $L_{Y'}$.

The theorem is proved.
$\blacksquare$\enddemo

\proclaim{Theorem 4.2}
Let $Y$ be a regular expression of order $\le n$ and weight $0$ (i.e
commuting with $h_i$ for all $i$),
and let $k$ be a complex
number ($k\notin \Bbb Q$). Then:

1) There exists a unique
differential operator $D_Y(k)\in DO_N$ such that
for all $\lambda$ for which $M_{\l,k}$ is irreducible
$$
<\Phi_0Y>=D_Y(k)<\Phi_0>,\tag 4.9
$$
satisfying the following conditions:

2) The operator $D_Y(k)$ is of order $\le n$ as a differential
operator;

3) The operator $D_Y(k)$ is a polynomial in $k$ of
degree $\le n$.

4) If $Y$ is a standard monomial and $\sdeg(Y)=s$ then
$D_Y=q^s\tilde D_Y$, where $\tilde D_Y\in \Cal DO_N$.
\endproclaim

\demo{Proof} First assume that $Y$ is a finite linear combination of
standard monomials. Since $Y$ is of weight $0$, the operator $L_Y$ preserves
weight in $\Vm$. Let $D_Y$ denote the restriction of $L_Y$ to the zero
weight subspace in $\Vm$. Obviously, $D_Y\in DO_N$, and it satisfies
properties 2,3,4. In particular, because it satisfies property 4,
the definition of $D_Y$ can be extended to infinite regular
expressions: if $Y=\sum_{m=1}^{\infty}b_mY_m$, where $Y_m$ are
standard monomials, then we can set
$$
D_Y=\sum_{m=1}^{\infty}b_mD_{Y_m},\tag 4.10
$$
and this series will be convergent as a formal series in $q$ since
any fixed power of $q$ occurs in only a finite number of its terms.

It remains to prove that $D_Y$ satisfying (4.9) is unique for any $Y$.
Suppose $D_Y^{(1)}$ and $D_Y^{(2)}$ both satisfy (4.9), and let
$D=D_Y^{(1)}-D_Y^{(2)}$. Then $D<\Phi>_{\lambda,k}=0$ for all $\l$.
Assume that $D\ne 0$.
Let $D_mq^m$, $D_m\in \Cal{DO}_N$,
be the leading term in the $q$-expansion of $D$. Since
$<\Phi>_{\l,k}\mid_{q=0}=\prod_{j=1}^Nz_j^{\lambda_j-k/N}$, we have:
$D_m(\prod_jz_j^{\l_j})=0$ for generic complex $\l_1,...,\l_N$. This implies
that $D_m=0$, which contradicts the assumption that $D_m$ is the
leading coefficient. Therefore, $D=0$, and hence
$D_Y^{(1)}=D_Y^{(2)}$.
$\blacksquare$\enddemo

Now we are in a position to define the mapping $\chi$. From now on the
notation $A^0$ will mean ``the zero weight part of $A$
(where $A$ is a subquotient of $\Uloc$). First we define
the linear mapping $\tilde\chi:\Uloc^0\to DO_N$ which acts according to the
formula:
$$
\tilde\chi(Y)=D_Y(-N).\tag 4.11
$$
It is easy to see that $D_{cY}=kD_Y$, which implies that $\tilde\chi$
kills the ideal $(c+N)\Uloc^0$, and therefore
descends to a mapping
$$
\chi:\Uloc^0/(c+N)\Uloc^0\to DO_N.\tag 4.12
$$
This mapping will be the main tool in our proof of the integrability
theorem.

For brevity we denote the algebra
$\Uloc/(c+N)\Uloc$ by $U_{\text{crit}}$. This notation is suggested
by the physical terminology: representations of $\hatglN$ with $c=-N$
in which $\Ucr$ acts are called critical level representations.

The mapping $\chi$ has the following important property:

\proclaim{Lemma 4.3} Let $C\in\Ucr$ be a central element: $[C,X]=0$
for any $X\in\Ucr$. Then
$$
\chi(CY)=\chi(Y)\chi(C),\ Y\in\Ucr^0 .\tag 4.13
$$
\endproclaim

\demo{Proof}
Let $\hat Y$, $\hat C$ be representatives of $Y,C$ in $\Uloc^0$.
Then for any $X\in\hatglN$ $[\hat C,X]=(c+N)X'$, $X'\in\Uloc^0$.
Therefore, it follows from the construction of the operators $D_Y$
(cf. proofs of Theorems 4.1, 4.2) that
$$
\gather
D_{\hat C\hat Y}<\Phi>=<\Phi\hat C\hat Y>=D_{\hat Y}<\Phi\hat C>+(k+N)E<\Phi>=
\\ D_{\hat Y}D_{\hat C}<\Phi>+(k+N)E<\Phi>,\ E\in DO_N,\tag 4.14
\endgather
$$
i.e. (by the uniqueness of $D_Y$)
$$
D_{\hat C\hat Y}(k)=D_{\hat Y}D_{\hat C}(k)+(k+N)E(k).\tag 4.15
$$
Therefore,
$$
D_{\hat C\hat Y}(-N)=D_{\hat Y}D_{\hat C}(-N),\tag 4.16
$$
which implies (4.13), Q.E.D.
$\blacksquare$\enddemo

 Lemma 4.3 implies

\proclaim{Theorem 4.4} Let $Z$ be the center of $\Ucr$. Then
the linear mapping $\chi:Z\to DO_N$ is a homomorphism of
algebras, and hence $\chi(Z)$ is a commutative subalgebra in
$DO_N$.
\endproclaim

\vskip .1in
\heading
{\bf 5. Sugawara operators and the integrability theorem}
\endheading
\vskip .1in

It is well known that the algebra $\Ucr$ has a big center $Z$.
This center is generated by the so-called Sugawara operators.
Explicit construction of these operators is very technical,
but for the proof of the integrability theorem we will need a
very limited amount of information about them.

Let $\xi_1,...,\xi_{N-1}$ is any basis of the Cartan subalgebra of
$\SLN$ orthonormal with respect to the inner product
$<A,B>=\text{Tr}AB$. We assume that $\xi_jt^m$, $m\ne 0$, are among the
basis vectors $X_i^{\pm}$. Also, if $a\in\SLN$, we denote $at^m$ as $a[m]$.

Sugawara operators for $\hatglN$ have orders $1,2,...,N$. We denote
the Sugawara operator of order $j$ by $\Omega_j$. The explicit
expressions for $\Omega_1$ and $\Omega_2$ are very simple:
$$
\gather
\Omega_1=\sum_{i=1}^Nh_i,\\
\Omega_2=\sum_{i=1}^Nh_i^2+\sum_{i\ne j}E_{ij}E_{ji}+\\
2\sum_{m=1}^{\infty}\sum_{i=1}^{N-1}\xi_i[-m]\xi_i[m]+
2\sum_{m=1}^{\infty}\sum_{i\ne j}E_{ij}[-m]E_{ji}[m]\tag 5.1
\endgather
$$

About higher order Sugawara operators we need to know the following:

\proclaim{Theorem 5.1} (\cite{*}, Proposition 3.3; see also \cite{*})
There exist elements
$T_j\in Z$, $3\le j\le N$, of order $j$,
which can be written in the following form:
$$
T_j=S_j+P_j(S_1,...,S_{j-1})+E_j,
\ S_m=\sum_{i=1}^Nh_i^m\tag 5.2
$$
where $P$ is a polynomial and $E_j$ contains only
standard monomials of positive subsidiary degree.
\endproclaim

\proclaim{Corollary 5.2} There exist degree zero elements $\Omega_j\in Z$
$3\le j\le N$, of order $j$,
which have the following form:
$$
\Omega_j=S_j+E_j^{\prime},\tag 5.3
$$
where $E_j^{\prime}$ contains only
standard monomials of positive subsidiary degree.
\endproclaim

\demo{Proof} Define the elements
$T_j^{(1)}$ of $Z$ as follows:
$$
T_j^{(1)}=T_j-P_j(T_1,...,T_{j-1}).\tag 5.4
$$
Then we have
$$
T_j^{(1)}=S_j+P_j^{(1)}(S_1,...,S_{j-2})+E_j^{(1)},\tag 5.5
$$
where $P_j^{(1)}$ is a polynomial, and all standard monomials in $E_j^{(1)}$
have positive subsidiary degree.

Next, define
$T_j^{(2)}\in Z$ by:
$$
T_j^{(2)}=T_j^{(1)}-P_j(T_1^{(1)},...,T_{j-1}^{(1)}).\tag 5.6
$$
Then we have
$$
T_j^{(2)}=S_j+P_j^{(1)}(S_1,...,S_{j-3})+E_j^{(2)},\tag 5.7
$$
where $P_j^{(2)}$ is a polynomial, and all standard monomials in $E_j^{(2)}$
have positive subsidiary degree.

Continuing this procedure, we will get to the $j-1$-th step, and
define $T_j^{(j-1)}$, which will have the decomposition
$S_j+E_j^{(j-1)}$. Therefore, we can set $\Omega_j=T_j^{(j-1)}$, Q.E.D.
$\blacksquare$\enddemo

Let us now assume that $z_j,q$ are complex numbers
such that $z_j\ne 0$, $0<|q|<1$,
and introduce new variables $x_j\in\Bbb C$ and $\tau\in\Bbb C^+$ such that
$e^{2\pi \text{i}x_j}=z_j$, $e^{2\pi\text{i}\tau/N}=q$.
For the proof of the integrability theorem we need one more
result:

\proclaim {Theorem 5.3}
$$
\chi(\Omega_2)=-\frac{\hat H}{4\pi^2}+\text{const},\tag 5.8
$$
where $\hat H$ is defined by the formula
$$
H=\sum_{j=1}^N\frac{\d^2}{\d x_j^2}-2\mu(\mu+1)\sum_{1\le i<j\le
N}\wp(x_i-x_j-\frac{j-i}{N}\tau|\tau).
\tag 5.9
$$
\endproclaim

\demo{Proof}
It is shown in \cite{EK, Proposition 4.2} that $\psi_{\lambda,k}$ satisfies the
following parabolic differential equation:
$$
-4\pi^2(k+N)q \frac{\d \psi_{\l,k}}{\d q}=(\hat H+c)\psi_{\l,k},\tag
5.10
$$
where $c$ is a constant. But we know that $(k+N)q \frac{\d
\psi_{\l,k}}{\d q}=<\Phi_0(k+N)\d>=<\Phi_0\Omega_2>=D_{\Omega_2}<\Phi_0>=
D_{\Omega_2}\psi_{\l,k}$, which implies that $D_{\Omega_2}=(4\pi^2)^{-1}\hat H
+\text{const}$, Q.E.D.
\enddemo

Now we are ready to prove the complete integrability theorem
for the elliptic case.

\proclaim {Theorem 5.4} (\cite{OP}) The Hamiltonian $H$ given by (1)
with $U=K\wp$ is completely integrable.
\endproclaim

\demo{Proof}. Let $\mu$ be such that $K=-2\mu(mu+1)$.
Consider the differential
operators $D_{\Omega_1}$,...,$D_{\Omega_N}$.
They are pairwise commutative, and, thanks to Corollary 5.2,
$$
D_{\Omega_j}=(2\pi
\i)^{-j}\sum_{l=1}^N\frac{\d^j}{\d x_j}+\text{lower order terms}
\tag 5.11
$$
Let us make a change of variable $\hat x_j=x_j+\frac{j}{N}\tau$,
and let $L_j$ be the image of $(2\pi \i)^jD_{\Omega_j}$
under this change of variable. Then $[L_j,L_m]=0$ for all $j,m$,
$L_2=H+\text{const}$ because of Theorem 3.3, and
$L_j=\sum_{l}\left(\frac{\d}{\d\hat x_l}\right)^j+\text{lower order
terms}$, which implies that $\{L_j\}$ are algebraically independent.
The theorem is proved.
\enddemo

\end